\documentclass[12pt]{article}
\usepackage{epsfig}

\input{tcilatex}

\begin{document}

\author{Kelly C. de Carvalho and T\^{a}nia Tom\'{e} \\
Instituto de F\'{\i}sica, Universidade de S\~{a}o Paulo \\
Caixa Postal 66318 \\
05315-970 S\~{a}o Paulo, SP, Brazil}
\title{Anisotropic probabilistic cellular automaton for a predator-prey
system}
\date{\today}
\maketitle

\begin{abstract}
We consider a probabilistic cellular automaton to analyze the stochastic
dynamics of a predator-prey system. The local rules are Markovian and are
based in the Lotka-Volterra model. The individuals of each species reside on
the sites of a lattice and interact with an unsymmetrical neighborhood. We
look for the effect of the space anisotropy in the characterization of the
oscillations of the species population densities. Our study of the
probabilistic cellular automaton is based on simple and pair mean-field
approximations and explicitly takes into account spatial anisotropy.
\end{abstract}

\newpage


\section{\ Introduction}

In the last years a particular great effort has been done in order to
understand the role of space given by a spatial structure and local
interactions in the characterization of the dynamics of competing biological
species systems \cite{tainaka,durrett,hastings,sat,tilman,sat1,durrett1,
droz,aguiar,kelly,albano,stauffer,szabo,kelly1,mobilia,tainaka2,arashiro,
kelly2}. In this context it has been studied irreversible stochastic lattice
models \cite{liggett, marro,ttmjo} with the purpose of mimic predator-prey
systems with Markovian local rules based in the Lotka-Volterra model \cite%
{lotka,volterra}. One of the problems that has been object of study is the
connection between the time oscillations of population densities and spatial
pattern distribution of the individuals of each species.

Here we study the coexistence of a two-species system by considering a
probabilistic cellular automaton (PCA) which is a modified version of the
automaton devised in \cite{kelly, kelly1,arashiro,kelly2}. The model, to be
called anisotropic predator-prey PCA possess local rules that are similar to
the ones of the cellular automaton proposed in \cite{nec} \ and was
introduced by us in order to explore the effect of spatial anisotropy in the
temporal oscillations.

We report dynamic mean-field approximations which take into account the
spatial dependence of densities and pair correlations of sites. In the next
section we present the model. In Sec. 2 we show the equations for the time
evolution of species densities and show the spatial dependence of these
equations. In Sec. 3 and 4 we perform dynamic simple and pair mean-field
approximations. Last section summarize the model, method and results.


\section{Model}

The physical space occupied by the species is represented by a regular
lattice of $N$ sites in which each site can be in one of three states. At
each site of the lattice we attach a stochastic variable $\eta_{i}$ that
takes the values $0$, $1$ and $2$ according whether the site $i$ is empty,
or occupied by a prey or occupied by a predator, respectively. The state of
the system can be represented by the vector $\eta=(\eta_1,\eta_2,\ldots,%
\eta_N)$. The time evolution equation for $P_{\ell}(\eta)$, the probability
of state $\eta$ at time $\ell$, is given by 
\begin{equation}
P^{\ell+1}(\eta)=\sum_{\sigma^{\prime}}W(\eta | \eta^{\prime}) P^{\ell
}(\eta^{\prime })  \label{mas}
\end{equation}
where the sum is over the $3^{N}$ configurations of the system and $W(\eta |
\eta^{\prime})$ is the transition probability from a state $\eta^{\prime}$
to state $\eta$, given that at the previous time step the system was in
state $\eta^{\prime }$. Since we are considering probabilistic cellular
automata, all the sites are updated simultaneously. In this case we have 
\begin{equation}
W(\eta | \eta^{\prime }) = \prod_{i=1}^{N} w_{i}(\eta_{i} | \eta^{\prime })
\label{master}
\end{equation}
where $w_{i}(\eta_{i} | \eta^{\prime })$ is the conditional transition
probability per site.


\subsection{The anisotropic predator-prey PCA}

\begin{figure}[tbp]
\centering
\epsfig{file=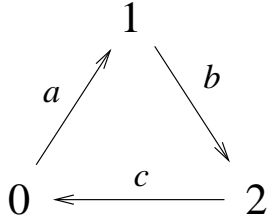,width=3.5cm}
\caption{Transitions of the predator-prey model. The three states are: prey
(1), predator (2) and empty (0). The allowed transitions obey the cyclic
order shown.}
\label{ppv}
\end{figure}

The stochastic rules, embodied in the transition rate $w_{i}(\eta
_{i}|\eta^{\prime })$, are set up in order that the allowed transitions
between states are only the ones that obey the cyclic order shown in Fig. %
\ref{ppv}. Prey can only born in empty sites; prey can give place to a
predator in a process where the prey dies and the predator is
instantaneously born; finally predator can die leaving an empty site. The
empty sites are places where prey can proliferate and can be seen as the
resource for prey surveillance. The death of predators complete this cycle,
reintegrating to the system the resources for prey.

The anisotropic predator-prey PCA has three parameters: $a$, the probability
of birth of prey, $b$, the probability of birth of predator and death of
prey, and $c$, the probability of predator death. Two of the process are
catalytic: the occupancy of a site by prey or by a predator is conditioned,
respectively, to the existence of prey or predator in the neighborhood of
the site. The third reaction, where predator dies, is spontaneous, that is,
it occurs, with probability $c$, independently of the neighbors of the site.
We assume that $a+b+c=1$ with $0\leq a,b,c\leq 1$.

The transition probabilities of the anisotropic predator-prey PCA are
described in what follows:

(a) If a site $i$ is empty, $\eta_{i}=0$, and if at north or east there is
at least one prey, then it can be occupied by a prey, $\eta_{i}=1$, in the
next time step, with a probability proportional to the parameter $a$ and to
the number of prey $n_{a}$ at north and east of site $i$.

(b) If a site is occupied by a prey, $\eta_{i}=1$, then the site has a
probability of being occupied by a new predator, $\eta _{i}=2$, in the next
time step if there are prey at north or east. In this process the prey dies
instantaneously. The transition probability is proportional to the parameter 
$b$ and the number of predators at north and east of site $i$.

(c) If site $i$ is occupied by a predator, $\eta_{i}=2$, it dies
spontaneously with probability $c$.

\begin{figure}[tbp]
\centering
\epsfig{file=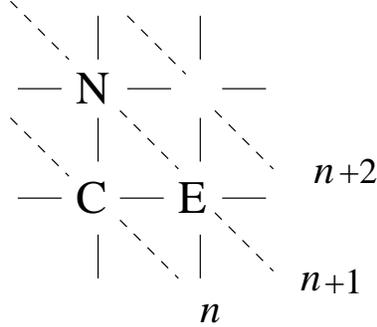,width=5cm}
\caption{A site (C) of the square lattice and its two nearest neighbor sites
to the east (E) and to the north (N). The layers, $n$, $n+1$, $n+2$, ... are
perpendicular to the southwest-northeast direction.}
\label{nec}
\end{figure}

The anisotropic cellular automaton is a variation of the automaton
introduced in \cite{kelly, kelly1,arashiro,kelly2}. Here each site of a
regular square lattice interacts with its first neighbors only at two
preferential directions. This anisotropic neighborhood consists of the
northern and eastern neighbors of each site as shown in Fig. \ref{nec}. The
set of transition probabilities per site is given by 
\begin{equation}
w_{i}(1|\eta )=\frac{a}2 n_1 \,\delta(\eta_{i},0) +(1-\frac{b}2 n_2)
\delta(\eta_{i},1),  \label{regra1}
\end{equation}
\begin{equation}
w_{i}(2|\eta )=\frac{b}2 n_2\,\delta(\eta_{i},1) +(1-c)\delta(\eta_i,2)
\label{regra2}
\end{equation}
and 
\begin{equation}
w_{i}(0 | \eta)=(1-\frac12 a n_1)\,\delta(\eta_{i},0) +c\delta(\eta_{i},2)
\label{regra3}
\end{equation}
where 
\begin{equation}
n_1=\sum_{k} \delta(\eta_{k},1) \qquad \mathrm{and}\qquad n_2=\sum_{k}
\delta(\eta_{k},2)
\end{equation}
and the sum is over the neighbor sites localized at east and north of site $%
i $ and correspond to the number of neighbors of site $i$ occupied by prey
individuals and predators individuals, respectively.

We have considered this probabilistic cellular automaton with the purpose of
verifying the effect of anisotropy in the properties of the time
oscillations of the predator-prey system. The rules considered are in some
sense inspired in the north-east-center (NEC) cellular automaton \cite{nec},
which also consider an unsymmetrical neighborhood of northern and eastern
sites.

The present stochastic dynamics predicts the existence of states, called
absorbing states, in which the system becomes trapped. Once the system has
entered such a state it cannot escape from it anymore remaining there
forever. There are two absorbing states. One of them is the empty lattice.
The other absorbing state is the one in which the lattice is full of prey.
This situation occurs if there are few predators and they become extinct.
The remaining prey will then reproduce without predation filling up the
whole lattice. The existence of absorbing stationary states is an evidence
of the irreversible character of the model or, in other words, of the lack
of detailed balance \cite{ttmjo}. However, the most interesting states, the
ones that we are concerned with in the present study, are the active states
characterized by the coexistence of prey and predators.


\subsection{Evolution equation for state functions}

The densities of prey, predator and empty sites are defined as 
\begin{equation}
P_i^{\ell}(1)=\langle \delta(\eta_i,1)\rangle_{\ell},
\end{equation}
\begin{equation}
P_i^{\ell}(2)=\langle \delta(\eta_i,2)\rangle_{\ell},
\end{equation}
\begin{equation}
P_i^{\ell}(0)=\langle \delta(\eta_i,0)\rangle_{\ell},  \label{def1}
\end{equation}
respectively. The lower index $i$ is used to denote the site and the upper
index $\ell$ stands for the time. The pair correlation of two neighbor sites 
$i$ and $j$ one being occupied by prey and the other by predator is defined
as 
\begin{equation}
P_{ij}^\ell( 
\begin{array}{cc}
1 & 2%
\end{array}%
) = \langle\delta(\eta_i,1)\delta(\eta_{j},2)\rangle_\ell.  \label{def2}
\end{equation}
In this definition the sites $i$ and $j$ are two neighboring horizontal
sites, the site $j$ being at the east of $i$. For two neighboring sites
placed vertically we use the notation 
\begin{equation}
P_{ik}^{\ell }\left( 
\begin{array}{c}
2 \\ 
1%
\end{array}%
\right) =\langle\delta(\eta_i,1)\delta(\eta_{j},2)\rangle_\ell,
\end{equation}
where $k$ denotes the neighbor of $i$ to the north.

The evolution equation for the densities can be obtained by using the rules
of the automaton and the evolution equation for the probability $%
P^{\ell}(\eta )$. Using equations (\ref{master}), (\ref{regra1}), (\ref%
{regra2}) and (\ref{regra3}), we obtain the time evolution equation for the
density of prey at site $i$, 
\begin{equation}
P_i^\prime(1)=\frac{a}{2} \left[ P_{ij}\left( 
\begin{array}{c}
1 \\ 
0%
\end{array}
\right)+P_{ik}( 
\begin{array}{cc}
0 & 1%
\end{array}
)\right]+P_i(1)-\frac{b}{2} \left[ P_{ij}\left( 
\begin{array}{c}
2 \\ 
1%
\end{array}
\right)+P_{ik}( 
\begin{array}{cc}
1 & 2%
\end{array}
)\right],  \label{dens1}
\end{equation}
where we have used prime and unprimed to denote quantities at time $\ell+1$
and $\ell$, respectively. Again, $P_{ij}( 
\begin{array}{c}
1 \\ 
0%
\end{array}
)$ denotes a pair correlation of a site $i$ which is empty and its neighbor
at the north $j$ which is occupied by a prey and $P_{ik}( 
\begin{array}{cc}
0 & 1%
\end{array}
)$ denotes the pair correlation of the site $i$ which is is empty and its
neighbor to the east $k$ which is occupied by a prey. Note that site $i$ is
the site to be updated. The time evolution equation for the density of
predators at site $i$ is given by 
\begin{equation}
P_i^\prime(2)=\frac{b}{2} \left[P_{ij}\left( 
\begin{array}{c}
2 \\ 
1%
\end{array}
\right)+P_{ik}( 
\begin{array}{cc}
1 & 2%
\end{array}
)\right]+(1-c)P_i(2).  \label{dens2}
\end{equation}

The evolution equations for correlations of two neighbor sites are given by
equations which involves clusters of two, three and four neighbors to the
north and east and they are too cumbersome. We will write them in the pair
approximation in the next subsection.

We will assume in the present analysis that the densities and the pair
correlations are homogeneous so that the correlations of sites placed
horizontally or vertically are the same. In other words the systems exhibits
a specular symmetry along the southwest-northeast line (see Fig. \ref{nec}).
Therefore, 
\begin{equation}
P_{ij}^{\ell }\left( 
\begin{array}{c}
1 \\ 
0%
\end{array}
\right)=P_{ik}^{\ell }( 
\begin{array}{cc}
0 & 1%
\end{array}
)  \label{nn1}
\end{equation}
and 
\begin{equation}
P_{ij}^{\ell }\left( 
\begin{array}{c}
2 \\ 
1%
\end{array}
\right)=P_{ik}^{\ell }( 
\begin{array}{cc}
1 & 2%
\end{array}
).
\end{equation}


\section{Simple mean-field approximation}

We will first analyze the equations (\ref{dens1}) and (\ref{dens2}) for the
densities of prey and predators by means of a simple mean field
approximation \cite{ttmjo,tome,tome1}. In this approximation we write the
probability of a given cluster of sites as the product of the probabilities
of each site. In our analysis we maintain the space dependence of each
probability. This is necessary since the rules that define the model are not
isotropic. We will use the notations 
\begin{equation}
x_n=P_n(1), \qquad y_n=P_n(2) \qquad z_n=P_n(0).
\end{equation}
where the index $n$ stands for a site at the $n$ layer. A layer is defined
as the sites belonging to a line perpendicular to the southwest-northeast
axis, as shown in Fig. \ref{nec}. Using this approximation and considering
equations (\ref{dens1}) and (\ref{dens2}) we obtain 
\begin{equation}
x_n^{\prime }=a z_n x_{n+1}+x_n - b x_n y_{n+1},  \label{cms1}
\end{equation}
and 
\begin{equation}
y_n^{\prime }=bx_n y_{n+1}+(1-c)y_n.  \label{cms2}
\end{equation}
where $z_n=1-x_n-y_n$. Due to this relation there is no need for the
equation for the density of empty sites.

The analysis of the above set of equations show that the stable solutions
are of two types: a prey absorbing state, where the all lattice is full of
prey; and an active solution with prey and predators densities constant in
time and space. So, we have just homogeneous and nonoscillating solutions
when we treat the anisotropic PCA by means of simple mean-field
approximation. At this level of mean-field approximation the space
anisotropy does not play any role in determining the kind of coexistence of
species.


\section{Pair-mean field approximation}

Now we consider the evolution equations for one-site correlations and two
site (pair) correlations. The evolution equations for the pair correlations
depend on clusters of three and four sites. Following the approach of the
pair approximation \cite{dickman,tome, tome1, tome4} we write the clusters
of three and four sites as products of pair correlations and one-site
correlations. However, we must be careful in doing this approximation
because we are maintaining the spatial anisotropy dependence of the one and
two site correlations. For example, the correlation of a site at layer $n$
in state $0$, a site at layer $n+1$ in state $0$ and a site at layer $n+2$
in state $1$, is written in the pair approximation as 
\begin{equation}
P_{n,n+1,n+2}(001)=\frac{P_{n,n+1}(00)P_{n+1,n+2}(01)}{P_{n+1}(0)}.
\label{appar}
\end{equation}
There are more complex clusters that appear in the evolution equation for
pair correlations. For example, the following cluster of four sites is
approximated by 
\begin{equation}
P_{n,n+1,n+2}\left( 
\begin{array}{ccc}
2 &  &  \\ 
0 & 0 & 1%
\end{array}
\right)=\frac{P_{n,n+1}(02)P_{n,n+1}(00)P_{n+1,n+2}(01)}{P_i(0)P_{n+1}(0)}.
\end{equation}
Observe that $P_{nn+1}(10)$ and $P_{nn+1}(01)$ are two independent variables.

Now we use the notation 
\begin{equation}
P_{n,n+1}(01) = u_n, \qquad P_{n,n+1}(12) = v_n, \qquad P_{n,n+1}(02) = w_n,
\end{equation}
\begin{equation}
P_{n,n+1}(10) = f_n, \qquad P_{n,n+1}(21) = g_n, \qquad P_{n,n+1}(20) = h_n,
\end{equation}
\begin{equation}
P_{n,n+1}(11) = r_n, \qquad P_{n,n+1}(22) = s_n, \qquad P_{n,n+1}(00) = q_n.
\end{equation}

\begin{figure}[tbp]
\centering
\epsfig{file=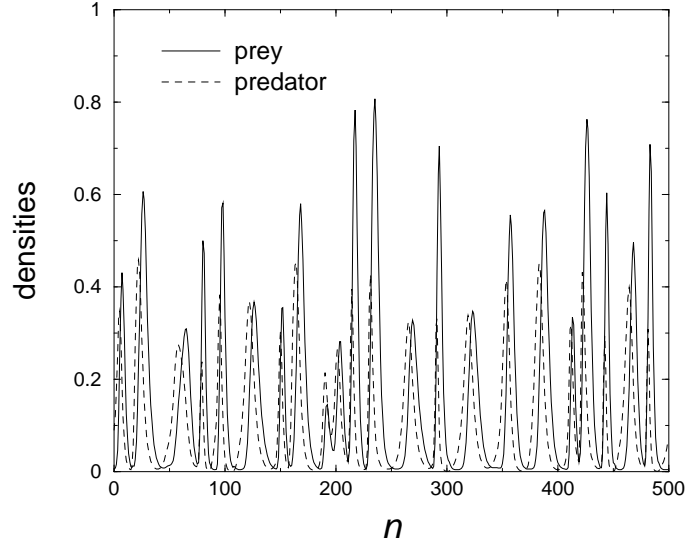,width=9cm}
\caption{Spatial oscillations of prey and predator along the
southwest-northeast direction.}
\label{oscespac}
\end{figure}

Using the pair approximation we derive the following equations for pair
correlations, 
\[
u_n^{\,\prime } = a\left(\frac{q_nu_n}{z_n}-a\frac{f_nq_nu_{n+1} }{z_nz_{n+1}%
}\right)+(1-\frac{a}{2})\left(u_n-b\frac{u_nv_{n+1}}{x_{n+1}}\right) 
\]
\[
-\frac{a}{2}\left(\frac{u_n^{2}}{z_n}-b\frac{u_n^{2}v_{n+1}}{ z_nx_{n+1}}%
\right)+ab\frac{h_nu_{n+1}}{z_{n+1}} 
\]
\begin{equation}
+c\left(g_n-b\frac{g_nv_{n+1}}{x_{n+1}}\right).  \label{eq_u}
\end{equation}
\[
v_n^{\,\prime } = \frac{ab}{2}\left(\frac{u_nv_{n+1}}{x_n}+\frac{
f_nu_nv_{n+1}}{z_nx_{n+1}}\right)+\frac{1}{2}a(1-c)\frac{f_nw_n}{x_n} 
\]
\[
+b\left(\frac{r_nv_{n+1}}{x_{n+1}}-\frac{b}{2}\frac{g_{n}r_{n}v_{n+1}}{
x_{n+1}^{2}}\right) 
\]
\begin{equation}
+(1-c)\left((1-\frac{b}{2})v_{n}-b\frac{v_{n+1}^{2}}{x_{n}}\right),
\label{eq_v}
\end{equation}
\[
w_{n}^{\,\prime} = b\left((1-\frac{a}{2})\frac{u_{n}v_{n+1}}{x_{n+1}}-\frac{a%
}{2} \frac{u_{n}^{2}v_{n+1}}{z_{n}x_{n+1}}\right) 
\]
\[
-a\frac{u_{n}^{2}v_{n+1}}{z_{n}x_{n+1}}+(1-c)\left(w_{n}-\frac{a}{2}\frac{
u_{n}w_{n+1}}{z_{n}}\right) 
\]
\begin{equation}
+bc\frac{g_{n}v_{n+1}}{x_{n+1}}+c(1-c)s_{n},  \label{eq_w}
\end{equation}
\[
f_{n}^{\,\prime } = a\left(\frac{u_{n}q_{n}}{z_{n}}-a\frac{u_{n}q_{n}u_{n+1}%
}{ z_{n}z_{n+1}}\right)-\frac{ac}{2}\frac{u_{n}w_{n}}{z_{n}} 
\]
\[
+f_{n}-a\frac{f_{n}u_{n}}{z_{n+1}}-\frac{b}{2}\left(\frac{v_{n}f_{n}}{x_{n}}%
-a \frac{v_{n}f_{n}u_{n+1}}{x_{n}z_{n+1}}\right) 
\]
\begin{equation}
+c\left((1-\frac{b}{2})v_{n}-\frac{b}{2}\frac{v_{n}^{2}}{x_{n}}\right).
\label{eq_f}
\end{equation}
\[
g_{n}^{\,\prime } = \frac{ab}{2}\frac{v_{n}f_{n}u_{n+1}}{x_{n}z_{n+1}}+\frac{%
1}{2 }b\left(\frac{v_{n}r_{n}}{x_{n}} -b\frac{v_{n}r_{n}v_{n+1}}{x_{n}x_{n+1}%
}\right) 
\]
\[
+a(1-c)\frac{h_{n}u_{n+1}}{z_{n+1}} 
\]
\begin{equation}
+(1-c)\left(g_{n}-b\frac{g_{n}v_{n+1}}{x_{n+1}}\right),  \label{eq_g}
\end{equation}
\[
h_{n}^{\,\prime } = \frac{b}{2}\left(\frac{v_{n}f_{n}}{x_{n}}-a\frac{
v_{n}f_{n}u_{n+1}}{x_{n}z_{n+1}}\right) 
\]
\[
-\frac{bc}{2}\left(v_{n}+\frac{v_{n}^{2}}{x_{n}}\right) +(1-c)\left(h_{n}-a%
\frac{ h_{n}u_{n+1}}{z_{n+1}}\right) 
\]
\begin{equation}
+c(1-c)s_{n}.  \label{eq_h}
\end{equation}

The equations for the densities are 
\begin{equation}
x_{n}^{\,\prime } = au_{n}+x_{n}-bv_{n,j},  \label{eq_x}
\end{equation}
and 
\begin{equation}
y_{n}^{\,\prime } = bv_{n}+(1-c)y_{n}.  \label{eq_y}
\end{equation}

Therefore, we get a set of eight equations that we have iterated to find the
solutions. We considered periodic boundary conditions and an initial
condition where half of the lattice with a set of densities and the other
half with another set of densities.

For $a=b$ and great values of $c$ the system attains a stationary state
which is the absorbing prey state. Decreasing $c$ there is a transition to
an active state, coexistence of species, and we found very interesting
solutions which are characterized by travelling waves of the densities of
each species. In Fig. \ref{oscespac} we show, for a particular set of the
parameters $a=b$ and $c=0.05$, the prey and predators oscillations as a
function of space and for a fixed instant of time. We can see that the
behavior of the space oscillations are very complex and very different from
a sinusoidal wave. 
\begin{figure}[tbp]
\centering
\epsfig{file=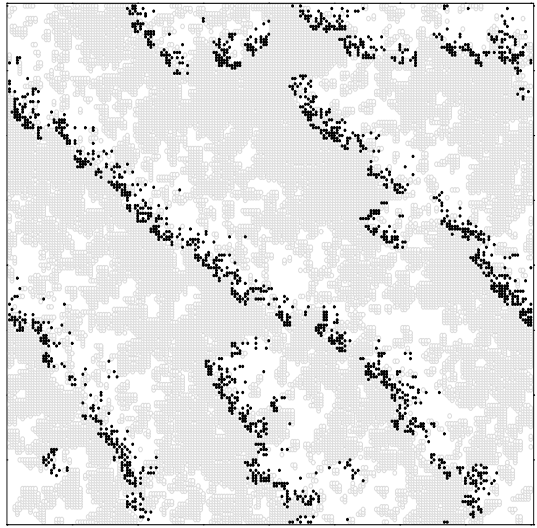,width=7cm,height=7cm}
\caption{Snapshot of a configuration obtained from Monte Carlo simulation of
the PCA in a square lattice with periodic boundary conditions. Predator and
prey individual are represented by black and grey points, respectively. The
fronts move from northeast to southwest.}
\label{snap}
\end{figure}

We remark that our assumption concerning the spatial dependence of densities
and pair correlations underlying our mean-field approach could not easily be
conceived a priori. It was set up by considering the type of local dynamics
with unsymmetrical rules and from the results of Monte Carlo simulations for
the present PCA \cite{kelly,tome2007}. To further clarify this point we show
in Fig. \ref{snap} a snapshot generated by simulation of the present PCA on
a square lattice where each site can be occupied by prey, predators or can
be empty. Periodic boundary conditions were used and the system evolved in
time according to the rules defined by Eqs. (\ref{regra1}), (\ref{regra2})
and (\ref{regra3}). A synchronous update was used. We see that the
distribution of individuals of each species is not homogeneous but exhibits
a pattern composed by layers of prey and predators. These layers are
displayed in a manner perpendicular to the southwest-northeast direction,
the direction along which the oscillations occur as can be seen in Fig. \ref%
{snap}. We have found that the spatial oscillations do not occur
independently of the time oscillations, since the spatial layers of prey and
predators are not static but are like fronts moving along the
southwest-northeast direction. The spatial pattern oscillations are
intimately associated to local time oscillations of the species \cite%
{kelly,tome2007}. These features were incorporated in the present mean-field
analysis of the model.


\section{Conclusions}

We have considered a predator-prey probabilistic cellular automaton with
anisotropic local stochastic rules. We studied this model by means of
dynamic mean-field approximation at two orders: simple mean-field
approximation and pair mean-field approximation. Due to the anisotropy these
approximations only can be performed if we consider the spatial dependence
of probabilities. The simple mean-field approximation for this automaton
just provides the prey absorbing state and active spatial homogeneous
solutions which are also constant in time. The pair mean-field approximation
gives much more rich results and show that the active states characterized
by time oscillations of species densities are inhomogeneous in space.
Therefore, we have spatiotemporal patterns of coexistence. This is in
accordance with previous Monte Carlo simulations \cite{kelly} where we have
found local time oscillations connected to inhomogeneous spatial
distributions of species individuals.



\begin{thebibliography}{99}
\bibitem{tainaka} K. Tainaka, Phys. Rev. Lett. \textbf{63}, 2688 (1989).

\bibitem{durrett} R. Durrett and S. Levin, Theor. Popul. Biol. \textbf{46},
363 (1994).

\bibitem{hastings} A. Hastings, \textit{Population biology: concepts and
models}, Springer, New York, 1997.

\bibitem{sat} J. Satulovsky and T. Tom\'{e}, Phys. Rev. E {bf 49}, 5073
(1994).

\bibitem{tilman} D. Tilman and P. Kareiva, \textit{Spatial ecology: the role
of space in population dynamics and interactions}, Priceton University
Press, Princeton, 1997.

\bibitem{sat1} J. Satulovsky and T. Tom\'{e}, J. Math. Biol. \textbf{35},
344 (1997).

\bibitem{durrett1} R. Durrett and S. Levin, J. Theor. Biol. \textbf{205},
201 (2000).

\bibitem{droz} T. Antal and M. Droz, Phys. Rev. E \textbf{63}, 056119 (2001).

\bibitem{aguiar} M. A. M. de Aguiar, H. Sayama, M. Baranger and Y. Bar-Yam,
Braz. J. Phys. \textbf{33}, 514 (2003).

\bibitem{kelly} K. C. de Carvalho and T. Tom\'{e}, Mod. Phys. Lett. B 
\textbf{18}, 873 (2004).

\bibitem{albano} A. F. Rozenfeld and E. V. Albano, Phys. Lett. A \textbf{332}%
, 361 (2004).

\bibitem{stauffer} D. Chowdhury and D. Stauffer, J. Biosc. \textbf{30}, 277
(2005).

\bibitem{szabo} G. Szab\'{o}, J. Phys. A \textbf{38}, 6689 (2005).

\bibitem{kelly1} K. C. de Carvalho and T. Tom\'{e}, Int. J. Mod. Phys. C 
\textbf{17}, 1647 (2006).

\bibitem{mobilia} M. Mobilia, I T Giorgiev and U. C. T\"{a}uber, Phys. Rev.
E \textbf{73}, 040903 (2006).

\bibitem{tainaka2} S. Morita S and K. Tainaka, Popul. Ecol. \textbf{48}, 99
(2006).

\bibitem{arashiro} E. Arashiro and T. Tom\'{e}, J. Phys. A \textbf{40}, 887
(2007)

\bibitem{kelly2} T. Tom\'{e} and K. C. de Carvalho, Stable oscillations in a
predator-prey system: a mean-field approach, arXiv:0704.0512v1 cond-mat
stat-mech.

\bibitem{liggett} T. M. Liggett, \textit{Interacting Particle Systems},
Springer, New York, 1985.

\bibitem{marro} J. Marro and R. Dickman, \textit{Nonequilibrium Phase
Transitions}, Cambridge University Press, Cambridge, 1999.

\bibitem{ttmjo} T. Tom\'{e} and M. J. de Oliveira, \textit{Din\^{a}mica Estoc%
\'{a}stica e Irreversibilidade}, Editora da Universidade de S\~{a}o Paulo, S%
\~{a}o Paulo, 2001.

\bibitem{lotka} A. Lotka, J. Am. Chem. Soc. \textbf{42}, 1595 (1920); Proc.
Nat. Acad. of Sciences USA \textbf{6}, 410 (1920); \textit{Elements of
Mathematical Biology}, Dover, New York (1924).

\bibitem{volterra} V. Volterra, \textit{Le\c{c}ons sur la Theorie Math\'{e}%
matique de la Lutte pour la Vie}, Gauthier-Villars, Paris, 1931.

\bibitem{nec} C. H. Bennett and G. Grinstein, Phys. Rev. Lett. \textbf{55},
657 (1985).

\bibitem{dickman} R. Dickman, Phys. Rev. A {bf 34}, 4246 (1986).

\bibitem{tome} T. Tom\'{e}, Physica A \textbf{212}, 99 (1994).

\bibitem{tome1} T. Tom\'{e} and J. R. Drugowich de Fel\'{\i}cio, Phys. Rev.
E \textbf{53}, 3976 (1996).

\bibitem{tome4} T. Tom\'{e}, E. Arashiro, J. R. Drugowich de Fel\'{\i}cio
and M. J. de Oliveira, Braz. J. Phys. \textbf{33}, 458 (2003).

\bibitem{tome2007} T. Tom\'e and K. C. de Carvalho, in preparation.
\end{thebibliography}
\end{document}